\documentclass[aps,prl,10pt,twocolumn,superscriptaddress]{revtex4-1}

\usepackage{times}
\usepackage{amsmath}
\usepackage{amsfonts}
\usepackage{graphicx}
\usepackage{mathtools}
\usepackage{hyperref}
\usepackage{braket}

\hypersetup{
colorlinks=true,
linkcolor=blue,
citecolor=blue,
filecolor=magenta,
urlcolor=cyan
}

\let\Re\relax

\DeclareMathOperator{\Tr}{Tr}
\DeclareMathOperator{\Re}{Re}

\newcommand{\papertitle}{Slow Growth of Out-of-time-order Correlators and Entanglement Entropy in Integrable Disordered Systems}
\newcommand{\authornames}{Max McGinley$^1$, Andreas Nunnenkamp$^1$, and Johannes Knolle$^{2,1}$}
\newcommand{\tcm}{Cavendish Laboratory, University of Cambridge, Cambridge CB3 0HE, United Kingdom}
\newcommand{\icl}{Blackett Laboratory, Imperial College London, London SW7 2AZ, United Kingdom}

\newcommand{\dif}{\mathrm{d}}

\begin{document}

\title{\papertitle}
\author{Max McGinley}
\affiliation{\tcm}
\author{Andreas Nunnenkamp}
\affiliation{\tcm}
\author{Johannes Knolle}
\affiliation{\icl}
\affiliation{\tcm}

\date{\today}

\begin{abstract}
	We investigate how information spreads in three paradigmatic one-dimensional models with spatial disorder. The models we consider are unitarily related to a system of free fermions and are thus manifestly integrable. We demonstrate that out-of-time-order correlators can spread slowly beyond the single-particle localization length, despite the absence of many-body interactions. This phenomenon is shown to be due to the nonlocal relationship between elementary excitations and the physical degrees of freedom. We argue that this non-locality becomes relevant for time-dependent correlation functions. In addition, a slow logarithmic-in-time growth of the entanglement entropy is observed following a quench from an unentangled initial state. We attribute this growth to the presence of strong zero modes, which gives rise to an exponential hierarchy of time scales upon ensemble averaging. Our work on disordered integrable systems complements the rich phenomenology of information spreading and we discuss broader implications for general systems with non-local correlations.
\end{abstract}

\maketitle

\bibliographystyle{apsrev4-1}

\emph{Introduction.---} The presence of spatial disorder in quantum systems can have profound effects on their static and dynamical properties, leading in particular to the phenomenon of localization \cite{Anderson1958, Kramer1993, Basko2006, Nandkishore2015, Altman2015, Nandkishore2015, Abanin2018}. Localized systems are often characterized by an absence of diffusion, and are therefore capable of retaining information about the initial state for arbitrarily long times.

Recently, there has been a surge of interest in studying localization in noninteracting and interacting systems \cite{Basko2006, Nandkishore2015, Altman2015, Nandkishore2015, Abanin2018}, termed Anderson (AL) and many-body localization (MBL), respectively. Although transport phenomena are the same in AL and MBL, the presence of interactions in MBL systems leads to a slow growth of entanglement entropy (EE) \cite{Znidaric2008, Bardarson2012, Serbyn2013}, indicating a propagation of information across the system, albeit at an exponentially slow rate.
In a similar light, it has recently been shown that out-of-time-order correlators (OTOCs) -- two-time correlation functions in which operators are not chronologically ordered -- are also capable of detecting this slow spread of information in MBL systems \cite{Huang2017, Swingle2017, Fan2017, Deng2017, Chen2017, Luitz2017}. Evidently, the presence of many-body interactions in localized systems has a drastic effect on the spreading of information, as witnessed by the EE and OTOCs. Excitingly, the realization of MBL systems in cold atom \cite{Schreiber2015, Choi2016} and trapped ion \cite{Smith2016} experiments, wherein the EE \cite{Daley2012,Islam2015} and OTOC \cite{Swingle2016, Zhu2016, Kaufman2016,Garttner2017,Garcia2017, Landsman2018} can be measured, allows for this slow information spreading to be directly observed \cite{Lukin2018}.

\begin{figure}
	\includegraphics[width=246pt]{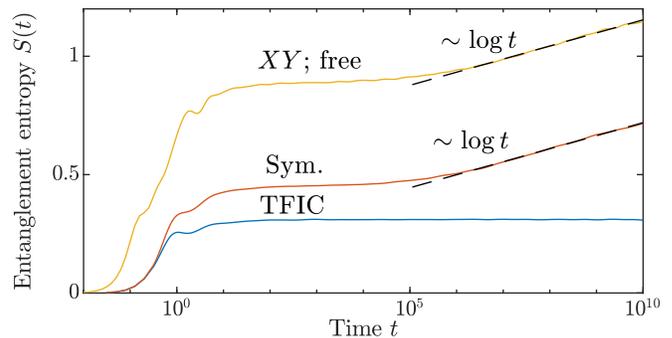}
	\caption{Growth of the second Renyi entropy after a quench under Hamiltonians \eqref{eqHamiltXY}, \eqref{eqHamiltFree}, and \eqref{eqHamiltSym}, and the transverse-field Ising chain (TFIC).
		We use clean $T_j = 1$ and disordered $R_j \in 2 + [-5,5]$. For each disorder realization, the initial state used is a random unentangled product state of fermion occupation or spin quantum numbers. The system size is $N=56$ in all cases, and the entropy is averaged over $M = 10^4$ disorder realizations. In systems \eqref{eqHamiltXY}, \eqref{eqHamiltFree}, and \eqref{eqHamiltSym} at late times, the entanglement entropy grows logarithmically despite the lack of many-body interactions usually associated with slow growth in MBL systems. The onset time of slow growth depends on the system size, whilst the final value of $S(t)$ as $t \rightarrow \infty$ is a constant of order 1, in contrast to MBL systems (see the discussion).
	}
	\label{figEntGrowth}
\end{figure}

\begin{figure*}
	\includegraphics[width=510pt]{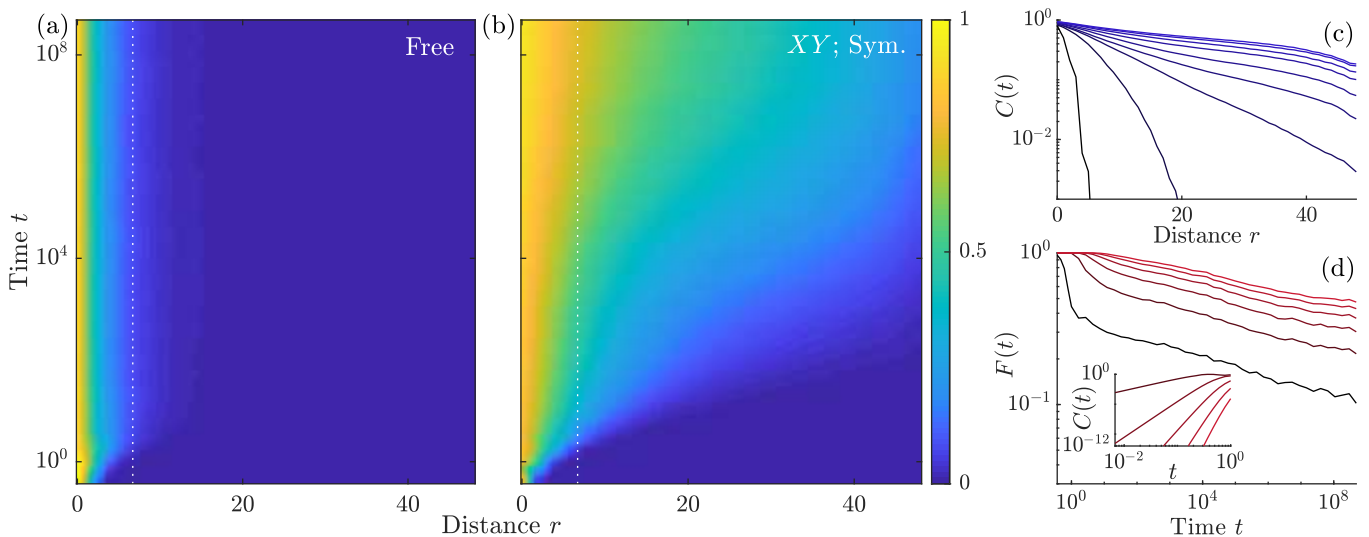}
	\caption{Dynamics of the out-of-time-order correlator $C(t)$ [Eq.\@ \eqref{eqOTOCDef} with $\hat{A}$ and $\hat{B}$ defined as a function of sites $j$ and $j+r$ in the main text] for integrable disordered models \eqref{eqHamiltXY}, \eqref{eqHamiltFree}, and \eqref{eqHamiltSym} with mean $\langle R \rangle = 1$ and width $W_R = 6$. All data is averaged over $M = 5 \times 10^3$ disorder realizations. The system size is $N = 100$ and we fix $j = 50$. Panel (a): color plot of $C(t)$ for the free fermion model as a function of time $t$ and operator distance $r$. Panel (b): equivalent color plot for the $XY$ and symmetric Kitaev models, which have identical OTOCs. The white dotted line indicates the single-particle localisation length. Panel (c): Data from (b) as a function of $r$ for fixed times $t$ varying from $t = 0.5$ (black) to $t = 5 \times 10^5$ (light blue), showing an exponential decay with time-dependent decay constant: $C \sim e^{-\lambda(t)|r|}$ -- this differs from the profile seen for typical MBL systems. Panel (d): Data from (b) [Plotted as $F(t) \equiv 1 - \Re C(t)$] as a function of $t$ for various distances $r$, varying from $r = 1$ (black) to $r = 23$ (red). Inset: short time behaviour of $C(t)$.
	}
	\label{figOTOCGrowth}
\end{figure*}

In this paper, focussing on EE and OTOCs, we study how information spreads in three disordered models whose Hamiltonians can be brought into free-fermion form and which, in that sense, are manifestly integrable. In all of our models, we observe slow dynamics in the EE which yields a logarithmic-in-time growth upon disorder averaging (Figure \ref{figEntGrowth}) -- we associate this growth with the presence of strong zero modes \cite{Fendley2016}. Furthermore, as our central result, we find that the OTOC slowly spreads beyond the single-particle localization length over long timescales (Figure \ref{figOTOCGrowth}) despite the lack of genuine many-body interactions. This we attribute to the nonlocal relationship between the physical and diagonal (free-fermion) bases, allowing nontrivial \textit{dynamical} correlations to appear which are not reflected in the static properties of eigenstates.

Although these signatures are generally associated with MBL phases, quantitative differences from typical MBL phenomenology are seen in the EE saturation value, which is order 1, and the profile of OTOC growth. Indeed the exact solvability of our models used here implies that slow OTOC and EE growth in localized systems is not always mediated by many-body interactions; thus this phenomenology cannot necessarily be used as signatures to distinguish AL and MBL systems. Our results highlight the role of these nonlocal correlations in non-equilibrium dynamics, and have broader implications for the diagnostics of localized phases.



\emph{Models.---} We study three disordered one-dimensional chains with open boundary conditions. Our first system of interest is the celebrated $XY$ spin chain with spatial disorder. The Hamiltonian is
\begin{align}
\hat{H}_{XY} = \sum_{j=1}^{N-1} T_j \hat{\sigma}^x_j \hat{\sigma}^x_{j+1} + R_j \hat{\sigma}^y_j \hat{\sigma}^y_{j+1}.
\label{eqHamiltXY}
\end{align}
The above Hamiltonian can be mapped to a 1D system of free fermions $\{\hat{f}_j^{(\dagger)}\}$ via the Jordan-Wigner (JW) transform $\hat{f}_j^{\dagger} = \hat{\sigma}_j^+\prod_{k<j}\hat{\sigma}_k^z$ \cite{Lieb1961}. The transformed system constitutes our second model, describing manifestly free fermions with anomalous terms
\begin{align}
\hat{H}_\text{free} =
\sum_{j=1}^{N-1} (T_j + R_j)\hat{f}_j^\dagger \hat{f}_{j+1} + (T_j - R_j)\hat{f}_j^\dagger \hat{f}_{j+1}^\dagger + \text{h.c.}
\label{eqHamiltFree}
\end{align}
This quadratic Hamiltonian, which is a disordered generalization of the Kitaev chain \cite{Kitaev2001}, can be efficiently diagonalized by a Bogoliubov transformation $\hat{a}_n = \sum_j u_{n,j} \hat{c}_j + v_{n,j} \hat{c}_j^\dagger$, such that $\hat{H}_\text{free} = \sum_n \epsilon_n \hat{a}_n^\dagger \hat{a}_n$ \cite{Lieb1961}.
The JW transform relates the eigenstates of (\ref{eqHamiltXY}) and (\ref{eqHamiltFree}) while preserving the spectrum.

Our third system can also be obtained through a JW transform, with the crucial difference that the $XY$ model is first rotated by $\pi/2$ into an `$XZ$' model. This yields a manifestly interacting fermionic Hamiltonian which is integrable, known as the symmetric interacting Kitaev chain \cite{Miao2017}
\begin{align}
\hat{H}_\text{sym} &=
\sum_{j=1}^{N-1} T_j (\hat{c}_j^\dagger \hat{c}_{j+1} + \hat{c}_j^\dagger \hat{c}_{j+1}^\dagger) +  \text{h.c.} \nonumber\\ &+ R_j(2\hat{c}^\dagger_j\hat{c}_j - 1)(2\hat{c}^\dagger_{j+1}\hat{c}_{j+1} - 1)
\label{eqHamiltSym}
\end{align}
\noindent with fermionic operators $\hat{c}_j^{(\dagger)}$. The system features hopping and $p$-wave pairing with equal amplitudes $T_j$, supplemented with nearest-neighbor density-density interactions.

After diagonalizing each system in the basis of quadratic Jordan-Wigner fermions, one can express the single-particle occupation numbers $\{\hat{a}_n^\dagger \hat{a}_n\}$ in the physical basis; this defines a collection of conserved quantities for each system. When either $T_j$ or $R_j$ are disordered, system \eqref{eqHamiltFree} exhibits Anderson localization \cite{Yu2013}, which makes each $\hat{a}_n^\dagger \hat{a}_n$ local in terms of $\hat{f}_j$ operators; we show that this locality also holds for the other two systems in the Supplemental Material (SM) \cite{SM}. Such an extensive set of local conserved quantities leads to the absence particle transport \cite{Huse2014}. This does not necessarily preclude information spreading -- e.g.~in MBL systems, interactions between the conserved quantities can lead to a slow growth of entanglement entropy \cite{Bardarson2012, Serbyn2013} and out-of-time-order correlators \cite{Huang2017, Swingle2017, Fan2017, Deng2017, Chen2017, Luitz2017}. Even so, since all our systems are spectrally equivalent to the Anderson insulator \eqref{eqHamiltFree}, such interactions are absent and we might expect that the EE and OTOC will quickly saturate to non-extensive values.

However, the presence of nonlocal `JW strings' in the transformations relating our systems plays an important role out of equilibrium. In systems \eqref{eqHamiltXY} and \eqref{eqHamiltSym}, the excitation operators $\hat{a}_n^\dagger$ which relate different eigenstates are highly nonlocal, 
unlike in a typical Anderson insulator.  We will see that the dynamics of these systems can unveil these nonlocal correlations which would otherwise cancel for eigenstates in equilibrium. The impact of JW strings on dynamical correlators for clean systems has been observed previously \cite{Rossini2009, *Rossini2010, Calabrese2011, *Calabrese2012, Lin2018}.

We use the Jordan-Wigner transforms to derive expressions for the EE and OTOC of all models in the SM \cite{SM}; these can be efficiently computed for large system sizes and long times. 

\emph{Entanglement entropy.---} In calculating the dynamics of the EE, our quench protocol is as follows: we begin in an unentangled product state of the relevant degrees of freedom (on-site fermion occupation numbers or spins $\hat{\sigma}^z$), and time evolve under a disordered Hamiltonian. In this paper we choose $T_j = 1$ and a uniform distribution for $R_j$, with mean $\langle R \rangle$ and width $W_R$. In each realization, we choose a random product state as the initial state, ensuring that the energy densities are equal on average. The time-evolved density matrix $\hat{\rho}(t)$ is partitioned into left and right halves of the system ($A$ and $B$, respectively), and the Renyi entropy $S^{(2)}(t) = -\ln \{[\Tr_B \hat{\rho}(t)]^2\}$ is calculated. We then average over $M = 10^4$ disorder realizations to obtain $\bar S^{(2)}(t)$.

Figure \ref{figEntGrowth} shows the disorder-averaged EE dynamics after a quench for each of the systems \eqref{eqHamiltXY}, \eqref{eqHamiltFree}, and \eqref{eqHamiltSym}. We also show results for the transverse-field Ising chain (TFIC) for reference, where the second term in \eqref{eqHamiltXY} is replaced by $R_j \hat{\sigma}_j^z$. The entanglement entropy first grows ballistically, before plateauing after a short time as expected for an AL system with a short localization length. However, we see that after a long time ($\sim 10^5$), the EE of the three systems of interest starts to slowly grow as $\log t$, in contrast to that of the TFIC which remains constant.
Such a logarithmic-in-time growth of EE is often associated with MBL phases \cite{Bardarson2012, Serbyn2013}, where interactions between conserved quantities lead to dephasing. However, since our systems are one-body reducible, they do not possess such interactions and so a different mechanism must be responsible. 

We attribute the unusual EE growth to the presence of nonlocal Majorana edge modes in systems \eqref{eqHamiltFree} and \eqref{eqHamiltSym}, which also manifest themselves as strong zero-modes of \eqref{eqHamiltXY} \cite{Fendley2016}, and to ensemble averaging. Except at criticality, these systems always possess edge modes \cite{Miao2017, McGinley2017} with an energy that is exponentially small in the system size. For a given realization of disordered $R_j$, we can estimate the energy of the zero mode as $|\log E_\text{maj}| \sim |\sum_j \log R_j|$ \cite{Fendley2012, McGinley2017}. The distribution of this energy scale is thus Gaussian in its \textit{logarithm}, and therefore the statistical ensemble of systems possesses an exponential hierarchy of timescales. Such a distribution of energy scales can in general lead to quantities which depend logarithmically on time \cite{Ariel2011}. Note that the TFIC does not possess zero modes for the parameters chosen, and thus does not exhibit this slow growth; similarly we have verified that the slow growth is absent in systems with periodic boundary conditions.

The above argument can be intuitively captured with a two-site fermionic toy model, described by four Majorana operators $\hat{\gamma}_{1,2}^{A,B}$. We construct a Hamiltonian which features one `edge mode' with annihilation operator $\hat{f}_e = \hat{\gamma}_1^A + i\hat{\gamma}_2^B$ and one `bulk mode' described by $\hat{f}_b = \hat{\gamma}_2^A + i\hat{\gamma}_1^B$. The Hamiltonian is $\hat H = E_\text{maj} \hat{f}_e^\dagger \hat{f}_e + E_b \hat{f}_b^\dagger \hat{f}_b$. We show in the SM \cite{SM} that if one averages the EE $S^{(2)}(t)$ for this model over the appropriate distribution of the edge mode energies, i.e.~$P(E_\text{maj}) \sim 1/E_\text{maj}$ for $E_- < E_\text{maj} < E_+$, then we obtain $\bar{S}^{(2)}(t) \propto \log t$ for times $E_+^{-1} < t < E_-^{-1}$. 

\emph{Out-of-time-order correlators.---} We now study the dynamics of OTOCs for our three systems. Specifically, we calculate the quantity (first proposed in Ref.\@ \cite{Larkin1969} and recently revived in \cite{Shenker2014, Kitaev2014})
\begin{align}
C(t) = \frac{1}{2}\left\langle [\hat{A},\hat{B}(t)]^\dagger [\hat{A},\hat{B}(t)] \right\rangle_\beta,
\label{eqOTOCDef}
\end{align}
\noindent where $\hat{A}$ and $\hat{B}$ are local Hermitian operators which commute and each square to 1, $\langle \cdot \rangle_\beta$ denotes a thermal expectation value at temperature $\beta^{-1}$. The above quantity contains the term $F(t) = \braket{\hat{A} \hat{B}(t) \hat{A} \hat{B}(t)}_\beta$ which features operators that are not time-ordered from right to left. Clearly, $C(t) = 1 - \Re F(t)$. The physical intuition behind this quantity is that, in a chaotic system, the operator support of $\hat{B}(t)$ will spread and eventually overlap with the support of $\hat{A}$, at which point $C(t)$ will become nonzero. Thus $C(t)$ measures operator spreading under the Hamiltonian of interest. The OTOC provides a way to understand how information spreads in localized systems. Logarithmic OTOC spreading has been proposed as a signature of MBL \cite{Huang2017, Swingle2017, Fan2017, Deng2017, Chen2017, Luitz2017}.

We compare how the OTOC develops in time for each of the systems \eqref{eqHamiltXY}, \eqref{eqHamiltFree}, and \eqref{eqHamiltSym}. We choose $\hat{A}_j$ and $\hat{B}_{j+r}$ to be the same local operator shifted by $r$ and will fix $j$ whilst varying $r$. We choose $\hat{A}_j$ to be $\hat{\sigma}^y_j$, $(2\hat{f}^\dagger_j \hat{f}_j - 1)$, and $(2\hat{c}^\dagger_j \hat{c}_j - 1)$ for the $XY$, free, and symmetric models, respectively.  We calculate the OTOC at infinite temperature by evaluating the operator in \eqref{eqOTOCDef} on a randomly selected eigenstate for each disorder realization. The OTOCs of models \eqref{eqHamiltXY} and \eqref{eqHamiltSym} can be shown to be identical. Importantly, the OTOC expression for these two cases features JW strings between sites $j$ and $j+r$.

The OTOC for our three models is plotted in Fig.~\ref{figOTOCGrowth}, which was calculated using the formula derived in the SM \cite{SM}. In the free fermion case, the OTOC spreads for a short time and then saturates at time $t \sim \mathcal{O}(1)$, as one would expect for an Anderson localized system. However, the presence of strings qualitatively changes the behavior of the OTOC for our systems \eqref{eqHamiltXY} and \eqref{eqHamiltSym}. As one of the central results of our work, we find that the OTOC does not saturate at short times, but spreads out. By plotting the OTOC at constant times, we see that $C(t)$ as a function of $r$ always decays exponentially with $r$, however the length scale of this decay increases with time beyond the static single-particle localization length, unlike in the free fermion case. For fixed distance the onset as a function of time as well as the approach to the long-time value appears to be power-law, similar to other integrable systems~\cite{Lin2018, Dora2017}.

\emph{Discussion.---} We have identified two features in the dynamics of our three disordered models, both of which lie beyond the physics expected for a typical Anderson insulator.

Firstly, we observed a slow logarithmic-in-time growth of the disorder-averaged bipartite entanglement entropy; we argued using a toy model that this was due to the presence of strong zero modes in our models.
The significance of the strong zero modes (as opposed to, e.g.\@ a ground state degeneracy due to spontaneous symmetry breaking) is twofold: it ensures that the \textit{entire} spectrum is nearly pairwise degenerate
; and it constitutes a mode whose wavefunction is delocalized between the two ends of the chain, such that it is picked up by the entanglement cut (see the Supplemental material \cite{SM} for details). We expect that this underlying mechanism
for slow entanglement dynamics also applies to non-integrable systems featuring strong zero modes, e.g.~parafermionic models \cite{Fendley2012}.

The limits of the energy distribution $E_+$ and $E_-$ determine the timescales when the logarithmic growth begins and ends. Their values depend on the Hamiltonian parameters as well as the system size. Away from criticality, the energies $E_{\pm}$ decrease for larger system sizes, leading to a later onset of logarithmic growth; this explains the late onset of slow growth in Fig.~\ref{figEntGrowth}. However, we expect this phenomenon to appear at earlier times in critical systems for arbitrarily large $N$, as well as in systems where edge modes appear at finitely separated topological domain walls. Moreover, the infinite time value of the entanglement entropy is expected to be a constant of order 1. Indeed our results appear to be consistent with previous studies of entanglement dynamics in the disordered $XX$ chain, i.e.~the critical version of the $XY$ model with $R_j = T_j$ \cite{DeChiara2006}.

Secondly, we observed a slow growth of the OTOC in models \eqref{eqHamiltXY} and \eqref{eqHamiltSym}. The profile of OTOC spreading we see is not typical for MBL or ergodic systems, where an `information front' emerges separating regions of $C(t) \approx 0$ and $C(t) \approx 1$ \cite{Huang2017, Swingle2017, Fan2017, Deng2017, Chen2017, Luitz2017}. This is in line with previous proofs of zero Lieb-Robinson velocities in related models \cite{Abdul2017}. However, the OTOC can reach appreciable values at spatial separations well beyond the single-particle localization length $\xi$
(dotted line in Fig.~\ref{figOTOCGrowth}), unlike one would expect for a typical Anderson localized system. 

Indeed, in the language of \cite{Huse2014}, the single-particle orbital occupations $\hat{a}_n^\dagger \hat{a}_n$ in the free system \eqref{eqHamiltFree} are a valid set of `l-bits': each forms a two-level system which can be defined locally in terms of the physical operators. However, the analogous quantities in the other models are not strictly l-bits, since the excitation operators $\hat{a}_n^\dagger$ are not local in the physical basis due to the JW strings. This subtlety does not affect the properties of static correlation functions, where $\braket{b|\hat{a}_n^\dagger |b}$ is necessarily zero if $\ket{b}$ is an eigenstate, but matrix elements between different eigenstates $\braket{b|\hat{a}_n^\dagger |c}$ are sensitive to this nonlocality, and such terms do appear in dynamical correlation functions.

Accordingly, let us express the OTOC $F(t)$ for an eigenstate $\ket{\Psi} = \ket{b}$ in a Lehmann representation, which gives (the states $\ket{b}$, $\ket{c}$, $\ket{d}$, $\ket{e}$ are all eigenstates of the Hamiltonian)
\begin{align}
F(t) &= \sum_{c,d,e} \braket{b|\hat{A}_j|c} \braket{c|\hat{B}_{j+r}|d} \braket{d|\hat{A}_j|e} \braket{e|\hat{B}_{j+r}|b} \nonumber\\ &\times \exp\left[i(E_b + E_d - E_c - E_e)t\right],
\label{eqOTOCDecomp}
\end{align}
and let us consider the long-time limit of the OTOC $F(\infty) \coloneqq \lim_{T \rightarrow \infty} \tfrac{1}{T}\int^T_0 \dif t' F(t')$ \cite{Huang2017b}. The terms with nontrivial dynamics $E_b + E_d - E_c - E_e \neq 0$ will oscillate and average to zero in the long-time limit, leading to a decay of $F(t)$ from its initial value $F(0) = 1$ (equivalently an increase of $C(t)$ from zero). We now discuss the criteria for nontrivial terms to have finite matrix elements, and hence for $C(t)$ to be nonzero.

Since all our systems are spectrally equivalent to the noninteracting Hamiltonian \eqref{eqHamiltFree} we can label energy eigenstates by their single-particle occupation numbers $\braket{b|\hat{a}_n^\dagger \hat{a}_n|b} \eqqcolon \eta_n^{(b)}$. Terms with nontrivial dynamics satisfy $\sum_n (\eta^{(b)}_n + \eta^{(d)}_n - \eta^{(c)}_n - \eta^{(e)}_n) \epsilon_n \neq 0$, for single-particle energies $\epsilon_n$. For a finite system, we assume that no two single-particle energies are commensurate, and so this quantity is only zero if $\tau^{(b)}_n + \tau^{(d)}_n - \tau^{(c)}_n - \tau^{(e)}_n = 0$ for all $n$ (i.e.~there are no `accidental' cancellations of the incommensurate $\epsilon_n$).

We therefore seek terms where $\eta^{(b)}_n + \eta^{(d)}_n - \eta^{(c)}_n - \eta^{(e)}_n \neq 0 $ for at least one $n$. From \eqref{eqOTOCDecomp}, one sees that if $\hat{A}_j$ has no overlap with the excitation operators $\hat{a}_n^\dagger$ and/or $\hat{a}_n$ (i.e.\@ $\hat{A}_j$ cannot cause a transition in the value of $\eta_n$), then we must have $\eta^{(b)}_n = \eta^{(c)}_n$ and $\eta^{(d)}_n = \eta^{(e)}_n$, so the term will be static; the same holds for $\hat{B}_{j+r}$. Therefore nonzero terms only arise when $\hat{A}_j$ and $\hat{B}_{j+r}$ can excite or de-excite the same single-particle orbital.

In generic AL systems, $\hat{A}_j$ will only be able to excite orbitals `near' site $j$, and similarly $\hat{B}_{j+r}$ acts only near site $(j+r)$. For sufficiently large $r$, it will not be possible for $\hat{A}_j$ and $\hat{B}_{j+r}$ to simultaneously act on the same orbital without incurring a factor of $e^{-r/\xi}$, where $\xi$ is the single-particle localization length, hence the OTOC will not spread beyond the length $\xi$. Additionally, for small $r$ only an $\mathcal{O}(1)$ number of orbitals can participate in the nonzero terms, and so the time at which the OTOC saturates to its long-time limit will also be $\mathcal{O}(1)$. This explains the fast saturation and spatial decay of the OTOC in Fig.~\ref{figOTOCGrowth}(a).

However, in systems \eqref{eqHamiltXY} and \eqref{eqHamiltSym}, the elementary excitations described by $\hat{a}_j^{(\dagger)}$ are nonlocal. This allows for $\hat{A}_j$ and $\hat{B}_{j+r}$ to act on the same orbital even when $r$ is large. Indeed, when one expresses the OTOC in the free-fermion $\hat{f}_j$ basis, non-cancelling JW strings appear \textit{between} sites $j$ and $(j+r)$, and so all orbitals in this range can participate in the contributing terms. This leads to the long-time spreading of OTOCs beyond the static single-particle localization length. The number of participating single-particle energies $\epsilon_n$ is $\mathcal{O}(r)$, and so the time taken to approach the long-time limit will also increase as $r$ increases, since there will be more nearly-cancelling terms with slow dynamics in \eqref{eqOTOCDecomp}. This explains the qualitative aspects of the OTOC growth seen in Fig.~\ref{figOTOCGrowth}(b). We expect that similar arguments hold for more general systems which also have nonlocal string correlations, leading to slow spatial growth of OTOCs.


Note that OTOC operators $\hat{A}_j$ and $\hat{B}_{j+r}$ with cancelling JW strings, e.g.\@ $\hat{A}_j = (\hat{c}_j - \hat{c}_j^\dagger)(\hat{c}_{j+1} + \hat{c}_{j+1}^\dagger)$ 
, would not be sensitive to the nonlocality of our systems, and we would see the same behavior as in Fig.~\ref{figOTOCGrowth}(a). This sensitivity to the choice of OTOC has been reported in the clean TFIC in Ref.\@ \cite{Lin2018}.

We note that 
fast oscillations in the OTOC for individual disorder realizations are expected even in the infinite-time limit due to the persistence of single-particle recurrences. The above arguments and the data shown in Fig.~\ref{figOTOCGrowth} characterize the long-time average over many disorder distributions, however the variance in the data is large, as one would expect.

%
\begin{acknowledgments}
	\emph{Acknowledgements.---}
	We thank Jens Bardarson, Markus Heyl, and Frank Pollmann for enlightening discussions, and Adam Smith for useful comments on the manuscript. M.~M.~holds an EPSRC studentship.
	A.~N.~holds a University Research Fellowship from the Royal Society and acknowledges support from the Winton Programme for the Physics of Sustainability.
\end{acknowledgments}

\bibliography{integrable_disorder}

\clearpage
\newpage
\appendix
	
\setcounter{figure}{0}
\makeatletter 
\renewcommand{\thefigure}{S\arabic{figure}}
	
\newcounter{defcounter}
\setcounter{defcounter}{0}
	
\newenvironment{myequation}
{%
\addtocounter{equation}{-1}
\refstepcounter{defcounter}
\renewcommand\theequation{S\thedefcounter}
\align
}
{%
\endalign
}

\pagenumbering{gobble}
	
\begin{widetext}
\begin{center}
{\fontsize{12}{12}\selectfont
\textbf{Supplemental Material for ``\papertitle''\\[5mm]}}
{\normalsize \authornames\\[1mm]}
{\fontsize{9}{9}\selectfont  
$^1$\textit{\tcm}\\
$^2$\textit{\icl}}
\end{center}
\normalsize
\end{widetext}

\subsection{Exact solution of models \eqref{eqHamiltXY} and \eqref{eqHamiltSym}}

In the main text, we stated that the three Hamiltonians \eqref{eqHamiltXY}, \eqref{eqHamiltFree}, and \eqref{eqHamiltSym} can all be related to each other via variations of Jordan-Wigner (JW) transformations. Here we explain the details of these transformations. For convenience, we will work with Majorana operators
\begin{myequation}
\hat{\gamma}_j^A &\equiv \hat{c}_j + \hat{c}_j^\dagger \nonumber\\
\hat{\gamma}_j^B &\equiv -i(\hat{c}_j - \hat{c}_j^\dagger) \nonumber\\
\hat{\lambda}_j^A &\equiv \hat{f}_j + \hat{f}_j^\dagger \nonumber\\
\hat{\lambda}_j^B &\equiv -i(\hat{f}_j - \hat{f}_j^\dagger).
\end{myequation}
In this basis, the Hamiltonians \eqref{eqHamiltFree} and \eqref{eqHamiltSym} can be written as
\begin{myequation}
\hat{H}_\text{free} &=
\sum_{j=1}^{N-1} -iT_j \hat{\lambda}^B_{j} \hat{\lambda}^A_{j+1} + iR_j \hat{\lambda}^A_{j} \hat{\lambda}^B_{j+1} \nonumber\\
\hat{H}_\text{sym} &=
\sum_{j=1}^{N-1} -iT_j \hat{\gamma}^B_{j} \hat{\gamma}^A_{j+1} - R_j \hat{\gamma}^A_{j}\hat{\gamma}^B_{j} \hat{\gamma}^A_{j+1} \hat{\gamma}^B_{j+1}.
\end{myequation}
In this basis, the transformations between our three systems can be written more concisely. 

Firstly, the $XY$ chain is solved exactly using a standard JW transform \cite{Lieb1961}. In the notation of Ref.~\cite{McGinley2017}, we write
\begin{myequation}
\hat{\lambda}_{2j-1}^A \equiv \hat{\phi}_{\text{I}, j}^A &= 
\left( \prod\nolimits_{k=1}^{2j-2} \hat{\sigma}_k^z \right)\hat{\sigma}^x_{2j-1}; \nonumber\displaybreak[0]\\
\hat{\lambda}_{2j}^A \equiv \hat{\phi}_{\text{II}, j}^A &=
 \left( \prod\nolimits_{k=1}^{2j-1} \hat{\sigma}_k^z \right)\hat{\sigma}^x_{2j}; \nonumber\displaybreak[0]\\
\hat{\lambda}_{2j-1}^B \equiv \hat{\phi}_{\text{I}, j}^B &=
 \left( \prod\nolimits_{k=1}^{2j-2} \hat{\sigma}_k^z \right)\hat{\sigma}^y_{2j-1}; \nonumber\displaybreak[0]\\
\hat{\lambda}_{2j}^B \equiv \hat{\phi}_{\text{II}, j}^B &=
\left( \prod\nolimits_{k=1}^{2j-1} \hat{\sigma}_k^z \right)\hat{\sigma}^y_{2j}.
\label{eqXYTrans}
\end{myequation}
If one expresses Hamiltonian \eqref{eqHamiltXY} in terms of these $\hat{\lambda}$ operators (which obey fermionic statistics due to the JW strings), one arrives at Eq.~\eqref{eqHamiltFree}. We have also defined two sets of $\hat{\phi}$ operators distinguished by a Roman numeral $\alpha = \mathrm{I},\, \mathrm{II}$, which reveals a decoupling into two independent subsystems. Specifically, \eqref{eqHamiltFree} is equivalent to
\begin{myequation}
\hat{H}_\text{free} &= \hat{H}_\text{free}^\text{I} + \hat{H}_\text{free}^\text{II} \nonumber\\ &= \sum_{j=1}^{N/2 - 1} \left[ -i t_{2j} \hat{\phi}_{\text{I},j+1}^A \hat{\phi}_{\text{I},j}^B + iU_{2j-1} \hat{\phi}_{\text{I},j}^A \hat{\phi}_{\text{I},j}^B \right] \nonumber\\&+ \sum_{j=1}^{N/2 - 1} \left[ i U_{2j} \hat{\phi}_{\text{II},j+1}^A \hat{\phi}_{\text{II},j}^B - it_{2j-1} \hat{\phi}_{\text{II},j}^A \hat{\phi}_{\text{II},j}^B \right].
\label{eqTransTF}
\end{myequation}
Each of the two subsystems is equivalent to a non-interacting Kitaev chain \cite{Kitaev2001} of length $N/2$, featuring a disordered chemical potential $\mu_j$ and equal hopping and superconducting pairing amplitudes $t_j = \Delta_j$. Using the methods of Lieb, Schultz, and Mattis \cite{Lieb1961}, one can obtain all eigenstates of the Hamiltonian in terms of the new $\hat \phi$ degrees of freedom.

The interacting symmetric Kitaev chain \eqref{eqHamiltSym} is solved using a less standard procedure. The observation of Miao \textit{et al.}~\cite{Miao2017} was that under an inverse JW transform, \eqref{eqHamiltSym} maps to a spin-1/2 chain with $\hat{\sigma}_j^x \hat{\sigma}_{j+1}^x$ and $\hat{\sigma}_j^z \hat{\sigma}_{j+1}^z$ couplings, which can be rotated to yield the $XY$ chain \eqref{eqHamiltXY}. We then employ the approach outlined above. The full mapping can be concisely written as the following unitary transformation
\begin{myequation}
\hat{\phi}_{\text{I}, j}^A &= 
\left( \prod\nolimits_{k\, \text{odd}}^{2j-3} i\hat{\gamma}_{k}^B \hat{\gamma}_{k+1}^A\right)\hat{\gamma}_{2j-1}^A; \nonumber\displaybreak[0]\\
\hat{\phi}_{\text{II}, j}^A &=
\left(\prod\nolimits_{k\, \text{odd}}^{2j-3} i\hat{\gamma}_{k}^A \hat{\gamma}_{k+1}^B\right)(i\hat{\gamma}_{2j-1}^A \hat{\gamma}_{2j}^A);
\nonumber\displaybreak[0]\\
\hat{\phi}_{\text{I}, j}^B &=
\left(\prod\nolimits_{k\, \text{odd}}^{2j-1} i\hat{\gamma}_{k}^B \hat{\gamma}_{k+1}^A\right)\hat{\gamma}_{2j}^B; \nonumber\displaybreak[0]\\
\hat{\phi}_{\text{II}, j}^B &=
\left(\prod\nolimits_{k\, \text{odd}}^{2j-3} i\hat{\gamma}_{k}^A \hat{\gamma}_{k+1}^B\right)(i\hat{\gamma}_{2j-1}^A \hat{\gamma}_{2j-1}^B).
\label{eqFullTrans}
\end{myequation}
This establishes the relationship between our three systems.

\section{Derivation of entanglement entropy and out-of-time-order correlators}

With the solutions given above, one can calculate relevant properties of systems \eqref{eqHamiltXY} and \eqref{eqHamiltSym}, both in and out of equilibrium. Here we provide derivations of the entanglement entropy and OTOC for both systems, which can be calculated to yield the results of Figures \ref{figEntGrowth} and \ref{figOTOCGrowth}.

\subsubsection{Renyi Entanglement Entropy}

Given a bipartition of a system into subregions $R$ and $S$, the reduced density matrix of a state $\ket{\Psi}$ is defined as $\hat{\rho}_R = \Tr_S \ket{\Psi}\bra{\Psi}$, where $\Tr_S$ denotes partial tracing over the degrees of freedom in $S$. The second Renyi entropy, which is a quantification of the entanglement between regions $R$ and $S$, is given by $S^{(2)} = -\log \Tr_R \hat \rho_R^2$ \cite{Eisert2010}.

By definition, the operator $\hat{\rho}_R$ contains all information required to calculate expectation values of operators $\hat{\mathcal{O}}_R$ that contain only degrees of freedom in $R$ via $\braket{\Psi|\hat{\mathcal{O}}_R|\Psi} = \Tr_R (\hat{\mathcal{O}}_R \hat{\rho}_R)$. Therefore, if $\Upsilon_R = \{\hat{\mathcal{O}}_i\}$ is a complete basis of many-body operators contained within $R$ which is orthonormal with respect to the Hilbert-Schmidt inner product $\Tr_R(\hat{\mathcal{O}}_i^\dagger \hat{\mathcal{O}}_j) = 2^{N_R} \delta_{i,j}$ then one can decompose the reduced density matrix as \cite{Fagotti2010}
\begin{myequation}
\hat{\rho}_R = \frac{1}{2^{N_R}}\sum_{\hat{\mathcal{O}}_i \in \Upsilon_R} \braket{\Psi|\hat{\mathcal{O}}_i|\Psi} \hat{\mathcal{O}}_i^\dagger.
\label{eqRhoDecomp}
\end{myequation}
We consider bipartitions that separate different spatial regions (as opposed to, e.g.~tracing out an internal degree of freedom). Thus a natural choice of basis $\Upsilon_R$ will be all possible products of spin operators $\{ \hat{\sigma}^\alpha_j\, |\, \alpha = x,y,z;\, j \in R\}$ or Majoranas $\{ \hat{\gamma}^\alpha_j\, |\, \alpha = A, B;\, j \in R\}$ for the $XY$ and Kitaev chains, respectively. A useful observation is that if $R$ is chosen to consist of the first $N_R$ sites, then the transformation \eqref{eqXYTrans} maps the first $N_R$ spin operators onto the first $N_R$ fermion operators, and vice versa (as well as preserving the Hilbert-Schmidt inner product) \cite{Latorre2004}. Therefore we can use the basis $\Upsilon_R = \{ \hat{\lambda}^\alpha_j\, |\, \alpha = A, B;\, j \in R\}$, even when $\hat{\lambda}$ are not the physical degrees of freedom. It is also possible to compute $\hat{\rho}_R$ for $R$ equal to a region between two cuts $j_-$ and $j_+$  by factoring out the string $\prod_{k=1}^{j_- -1} (i \hat \gamma^A_k \hat \gamma^B_k)$, but we will focus on the simpler case in the following.

We call a density matrix `Gaussian' if it can be expressed in the form $\hat{\rho}_R = \mathcal{Z}^{-1} \exp ( \hat{\lambda}_i M_{i,j} \hat{\lambda}_j)$ \cite{Peschel2003}, and similarly a state is Gaussian if its (reduced) density matrix is Gaussian. Gaussian density matrices are simple to work with because all expectation values can be evaluated using Wick's theorem.

We are interested in the Renyi entropy of a time-evolved state $\ket{\Psi(t)}$ which started as an unentangled product state $\ket{\Psi(0)}$ at $t=0$. In the free-fermion system \eqref{eqHamiltFree} (described by $\hat{\lambda}$ Majorana operators, defined in terms of the $\hat{f}^{(\dagger)}$ Dirac operators), we can choose a fermion vacuum $\hat{f}_j\ket{\Psi_\text{free}(0)} = 0$ which is indeed a Gaussian state. Similarly, for model \eqref{eqHamiltXY} we can start with a paramagnetic state $(\hat{\sigma}_j^z - 1)\ket{\Psi_{XY}(0)} = 0$, and for \eqref{eqHamiltSym} we use the state such that $\hat{c}_j \ket{\Psi_\text{Sym}(0)} = 0$.

One might expect that when expressed in terms of $\hat{\lambda}$ operators, these simple states are all Gaussian, as they can each be thought of as the ground state of some quadratic-in-$\hat{\lambda}$ Hamiltonian. However, as noted in \cite{Eisler2016}, JW transforms do not necessarily map Gaussian states to other Gaussian states. The initial state $\ket{\Psi_\text{Sym}(0)}$ is in fact a superposition of two Gaussian states (sometimes called a `cat state'). To see this, note that $\ket{\Psi_\text{Sym}(0)}$ can be thought of as the ground state of \eqref{eqHamiltSym} with $T_j = 0$ and $R_j = -1$. Because $\hat{H}_\text{sym}$ commutes with the operator $\hat{Z}_2^f = \prod_{k\, \text{odd}} i\hat{\gamma}_k^A \hat{\gamma}_{k+1}^B$ which exchanges occupied and unoccupied sites, the ground states one obtains are $\ket{P} = (\ket{\Psi} + \ket{\bar{\Psi}})/\sqrt{2}$ and $\ket{Q} = (\ket{\Psi} - \ket{\bar{\Psi}})/\sqrt{2}$ with $Z_2^f$-eigenvalues $+1$ and $-1$, respectively. (We use $\ket{\bar{\Psi}}$ to denote the state with opposite fermion occupation numbers.) Whilst $\ket{P}$ and $\ket{Q}$ are Gaussian, the superposition of them is not. The states are related by an operator $\ket{Q} = \hat{\mathcal{M}}\ket{P}$ with $\mathcal{\hat M} = \hat{\lambda}_1^B$. It is important to note that $\ket{P}$ and $\ket{Q}$ have opposite fermion number parities in the $\hat{\lambda}$ basis.

If we expand \eqref{eqRhoDecomp} into four terms
\begin{myequation}
\hat{\rho}_R &= \frac{1}{2}\left(\hat{\rho}_{PP} + \hat{\rho}_{PQ} + \hat{\rho}_{QP} + \hat{\rho}_{QQ}\right) \nonumber\\
&= \frac{1}{2^{N_A+1}}\sum_{\hat{\mathcal{O}}_i \in \Upsilon_R} \left[ \braket{P|\hat{\mathcal{O}}_i|P} + \braket{P|\hat{\mathcal{O}}_i|Q}\right.\nonumber\\
&+ \left.\braket{Q|\hat{\mathcal{O}}_i|P} + \braket{Q|\hat{\mathcal{O}}_i|Q}  \right] \hat{\mathcal{O}}_i^\dagger
\end{myequation}
then, as $\ket{P}$ and $\ket{Q}$ are representable by Gaussian states, the terms $\hat{\rho}_{PP}$ and $\hat{\rho}_{QQ}$ are Gaussian density matrices. Specifically, one can show that \cite{Fagotti2010}
\begin{myequation}
\hat{\rho}_{PP} &= \frac{1}{\mathcal{Z}_{P}}\exp\left(\frac{\hat{\lambda}_a W_{a,b}^{(P)} \hat{\lambda}_b }{4}\right),\nonumber\\
\text{where}\, W^{(P)} &= \tanh(\Gamma^{(P)}/2)\nonumber\\ \text{and}\, \mathcal{Z}_P &= \sqrt{\det \cosh(W^{(P)}/2)}.
\end{myequation}
$W^{(P)}$ is a $2N_R \times 2N_R$ antisymmetric matrix of $c$-numbers, and the matrix $\Gamma^{(P)}_{a,b}=\braket{P|\hat{\lambda}_a \hat{\lambda}_b|P} - \delta_{a,b}$ is the two-particle correlation matrix of the state $\ket{P}$, restricted to the indices $a,b \in \Upsilon_R$. An equivalent expression holds for $\hat{\rho}_{QQ}$.

The remaining two terms cannot be written in Gaussian form because they contain only terms with an odd number of $\hat{\lambda}$ operators (since $\ket{P}$ and $\ket{Q}$ have opposite fermion number parity). However, as $\braket{P|\hat{\mathcal{O}}_i|Q}$ can be written as $\braket{P|\hat{\mathcal{O}}_i\hat{\mathcal{M}}|P}$, we expect that one should still be able to use some version of Wick's theorem to calculate all expectation values. We therefore suggest a form for  $\hat{\rho}_{PQ}$
\begin{myequation}
\hat{\rho}_{PQ} &= \exp \left(\frac{\hat{\lambda}_a W_{a,b}^{(PQ)} \hat{\lambda}_b }{4}\right)(u_k^{(PQ)} \hat{\lambda}_k)
\label{eqRhoPQ}
\end{myequation}
\noindent where $u_k^{(PQ)}$ and $W_{a,b}^{(PQ)}$ are $c$-numbers and all the labels of $\hat{\phi}$ operators are contained within a single index $k$. The above form satisfies the requirements of $\hat{\rho}_{PQ}$ in that it only contains odd numbers of $\hat{\phi}$ operators and generates multiparticle moments via Wick's theorem, due to the exponential factor. Whilst we have no direct proof of \eqref{eqRhoPQ}, we are able to show that it generates all the correct expectation values between $\ket{P}$ and $\ket{Q}$, given an appropriate choice of $u_k^{(PQ)}$ and $W_{a,b}^{(PQ)}$. One finds that the appropriate choice is
\begin{myequation}
u_k^{(PQ)} &= \frac{1}{\mathcal{Z}_P}\sum_{\beta = 1}^N(1 - \Gamma^{(P)})_{k,\beta} m_\beta \nonumber\\
W_{a,b}^{(PQ)} &= \tanh\left(\frac{\Gamma^{(P)}}{2}\right)_{a,b},
\end{myequation}
\noindent where $m_\beta$ defines the Majorana operator $\hat{\mathcal{M}}$ through $\hat{\mathcal{M}} = m_\beta \hat{\lambda}_\beta$. Here we distinguish Greek letters which run over all the indices in the system $\beta = 1,\ldots, N$ from Latin letters which run over only the indices contained in $\Upsilon_R$ $a = 1,\ldots,|R|$. Finally $\hat{\rho}_{QP}$ can be determined as it is the Hermitian conjugate of $\hat{\rho}_{PQ}$.

Now that we have written $\hat{\rho}_R$ as a sum of four operators which are either in Gaussian or modified-Gaussian form, we can compute the second Renyi entropy, which involves $\Tr_R \hat{\rho}_R^2$. Because the trace of the product of any non-zero number of inequivalent Majorana operators is zero, we get separate contributions from the Gaussian parts $\hat{\rho}_{PP} + \hat{\rho}_{QQ}$ and the modified Gaussian parts $\hat{\rho}_{PQ} + \hat{\rho}_{QP}$. Using the algebra developed by Fagotti and Calabrese \cite{Fagotti2010}, we can calculate this first part using the `product rule'
\begin{myequation}
\Tr_R \left(\hat{\rho}_{PP}\hat{\rho}_{QQ}\right) &= \{\Gamma^{(P)}, \Gamma^{(Q)}\} \nonumber\\ &\coloneqq \sqrt{\det \left| \frac{1 + \Gamma^{(P)}\Gamma^{(Q)}}{2} \right|}
\end{myequation}
\noindent with two more similar terms required. The second part involves the trace of two operators of the form \eqref{eqRhoPQ}. One can commute the linear-$\hat{\lambda}$ part through the exponential, and then combine the exponentials using Equation (41) in Ref.~\onlinecite{Fagotti2010}. We are then left with the trace of a single quadratic exponential with a sum of fermion bilinears, which is just a two-particle expectation value. We compute these expectation values and combine all the terms together, yielding
\begin{myequation}
4e^{-{S^{(2)}}} &= 4\Tr_R \hat{\rho}_R^2 \nonumber\\
&= \{\Gamma^{(P)}, \Gamma^{(P)}\}\left(1 + p^T (1 + (\Gamma^{(P)})^2)^{-1} p\right) \nonumber\\
&+ 2\{\Gamma^{(P)}, \Gamma^{(Q)}\}\left(1 + p^T (1 + \Gamma^{(Q)}\Gamma^{(P)})^{-1} q\right) \nonumber\\
&+ \{\Gamma^{(Q)}, \Gamma^{(Q)}\}\left(1 + q^T (1 + (\Gamma^{(Q)})^2)^{-1} q\right),
\label{eqRenyi}
\end{myequation}
\noindent where we use the shorthand $p_k = \sum_{\beta=1}^N(1-\Gamma^{(P)})_{k,\beta}m_\beta$ and $q_k = \sum_{\beta=1}^N(1-\Gamma^{(Q)})_{k,\beta}m_\beta$. One can apply the above to the time-evolved state $\ket{\Psi(t)} = e^{-iHt}\ket{\Psi(0)}$, which just involves calculating the time-dependent correlation matrices $\Gamma^{(P,Q)}(t)$ and the time-evolved vector $m_\beta(t)$. We finally arrive at the desired time-dependent Renyi entanglement entropy.

\subsubsection{OTOC}

To calculate an OTOC \eqref{eqOTOCDef} for an eigenstate of the integrable Hamiltonian \eqref{eqHamiltSym}, we make use of the fact that $\hat{H}$ is quadratic in $\hat{\lambda}$ operators. This means that time evolution can be easily performed, and that we can exploit Wick's theorem to calculate the multiparticle expectation values required. However, if we attempt to use the usual Wick theorem for time-ordered expectation values, we will inadvertently calculate the correlator with all operators in the usual time order. To overcome this, we must generalize the time-ordering scheme to the so-called augmented Keldysh convention developed by Aleiner \textit{et al.}~\cite{Aleiner2016}.

The method involves redefining the time-ordering operator $\mathcal{T}$ in a way that allows operators to appear in the desired order seen in \eqref{eqOTOCDef}. The $\hat{\lambda}$ degrees of freedom are replicated four times (as opposed to two times in the standard Keldysh convention), i.e.~we assign a label $\mu = 1,2,3,4$ to the operator which does not affect how the operator acts on the wavefunction, but does change the ordering of the operators under the new augmented time-ordering operator $\mathcal{T}_{\mathcal{C}_K}$. We can understand this as time evolution along an augmented contour $\mathcal{C}_K$ -- see Ref.~\onlinecite{Aleiner2016} for details. Specifically, we define this new time-ordering as
\begin{myequation}
\mathcal{T}_{\mathcal{C}_K} \hat{\phi}_j^\mu(t_1) \hat{\phi}_k^\nu(t_2) = \begin{dcases*}
\hat{\phi}_j^\mu(t_1) \hat{\phi}_k^\nu(t_2) & $\mu > \nu$; \\
-\hat{\phi}_k^\nu(t_2) \hat{\phi}_j^\mu(t_1) & $\mu < \nu$; \\
\mathcal{T}_\text{ord} \hat{\phi}_j^\mu(t_1) \hat{\phi}_k^\nu(t_2) & $\mu = \nu$ is even; \\
\mathcal{T}_\text{rev} \hat{\phi}_j^\mu(t_1) \hat{\phi}_k^\nu(t_2) & $\mu = \nu$ is odd.
\end{dcases*}
\end{myequation}
Here, $\mathcal{T}_\text{ord}$ is the ordinary time-ordering operator which ensures the greater time appears on the left, and $\mathcal{T}_\text{rev}$ is the reverse time-ordering operator which ensures the greater time appears on the right. In words, the contour index $\mu$ takes precedence in time ordering, and otherwise we time-order in the forwards or backwards direction depending on the index. We now define the augmented Green's function matrix
\begin{myequation}
\mathcal{G}^{\mu, \nu}_{j,k}(t_1, t_2) = \braket{\Psi|\mathcal{T}_{\mathcal{C}_K} \hat{\phi}_j^\mu(t_1) \hat{\phi}_k^\nu(t_2)|\Psi}.
\label{eqGreens}
\end{myequation}

With this formalism developed, it is then simple to calculate OTOCs. After expressing the operators $\hat{A}_j$ and $\hat{B}_{j+r}$ in terms of $\hat{\lambda}$ operators, we assign contour indices to the operators according to their place in the OTOC. We write
\begin{myequation}
F(t) = \braket{\Psi|\mathcal{T}_{\mathcal{C}_K}\hat{A}_j^4 \hat{B}_{j+r}^3 (t) (\hat{A}_j^2)^\dagger (\hat{B}_{j+r}^1 (t))^\dagger|\Psi}
\label{eqOTOCOrdered}
\end{myequation}
\noindent where the superscripts are contour indices. We can now freely use Wick's theorem to compute the above as products of appropriate components of the Green's function matrix \eqref{eqGreens}, and the redefined time-ordering operator will ensure the result corresponds to the desired order \eqref{eqOTOCDef}. As is often the case when applying Wick's theorem to Majorana operators, we can make use of the `Pfaffian trick' \cite{Bravyi2012}. If one considers $\mathcal{G}^{\mu, \nu}_{j,k}(t_1, t_2)$ as a matrix with rows labelled by $(\mu, j)$ and columns labelled by $(\nu, k)$, then the OTOC is equal to the Pfaffian of the submatrix that only contains rows and columns corresponding to the $\hat{\lambda}$ operators which appear in \eqref{eqOTOCOrdered} after writing the right hand side in terms of $\hat{\lambda}$s. In practice, following Ref.~\cite{Lin2018}, to evaluate the Pfaffian we calculate the determinant and take its square root, choosing the sign such that the derivative is continuous in space (i.e.~for consecutive values of $r$).

The local unitary operators $\hat{A}_j$ and $\hat{B}_{j+r}$ which we choose in the main text are $\hat{A}_j = (2\hat{n}_j - 1) \equiv i\hat{\gamma}_j^A \hat{\gamma}_j^B$, and similarly for $\hat{B}_{j+r}$. In terms of $\hat{\lambda}$ fermions, this is
\begin{myequation}
i\hat{\gamma}^A_j \hat{\gamma}^B_j = \left(\prod_{k=1}^{j-1} (i\hat{\lambda}_{k}^A \hat{\lambda}_{k}^B)\right)\hat{\lambda}_{j}^B.
\label{eqTransformedOccupation}
\end{myequation}
The above expression features an operator string from site 1 to $j$, and is thus highly non-local in the rotated basis. Indeed when we evaluate \eqref{eqOTOCOrdered}, we will have to include all $\hat{\lambda}$ operators between sites $j$ and $(j+r)$, in contrast to the OTOC evaluated for a non-interacting system, which only features operators near $j$ and $j+r$ separately. Whilst the JW strings cancel in equilibrium (i.e.~at $t=0$, where they commute through and square to unity), their dynamics plays an important role out of equilibrium, and is responsible for the OTOC growth seen in Fig.~\ref{figOTOCGrowth}.

We note in passing that, although there exists an exact equality between the Renyi entropy and a particular sum of OTOCs \cite{Fan2017}, this is not reflected in the results of Figures \ref{figEntGrowth} and \ref{figOTOCGrowth}, since the conditions for such a relationship to hold are not fulfilled by our protocols. In particular, the requirement that $\hat{B}_j$ can be written as $\hat{O}\hat{O}^\dagger$ would necessitate the cancellation of JW strings in the OTOC.

\section{Toy model}

Here we explicitly calculate the entanglement properties of the two-site toy model introduced in the main text, which captures the influence of the edge mode on entanglement dynamics. Using four Majorana operators $\hat{\gamma}_{1,2}^{A,B}$, we construct a Hamiltonian which resembles the true system: we include one `edge mode' with an annihilation operator $\hat{f}_e = \hat{\gamma}_1^A + i\hat{\gamma}_2^B$ and one `bulk mode' described by $\hat{f}_b = \hat{\gamma}_2^A + i\hat{\gamma}_1^B$. The Hamiltonian is

\begin{myequation}
H = E_\text{maj} \hat{f}_e^\dagger \hat{f}_e + E_b \hat{f}_b^\dagger \hat{f}_b.
\end{myequation}

If we start with an initial unentangled product state with occupations $\eta_{1,2} = \pm 1$ on sites 1 and 2, then we can calculate the entanglement entropy between the two sites

\begin{myequation}
S^{(2)}(t) = \log 2 - \log\left(1 + \cos^2 \left[(E_\text{maj} + \eta_1 \eta_2 E_b)t\right]\right).
\end{myequation} 

\noindent For concreteness we choose $\eta_1 = -\eta_2 = 1$.

The energies $E_b$ and $E_\text{maj}$ are random and should be drawn from appropriate distributions $P_b(E_b)$ and $P_\text{maj}(E_\text{maj})$. Whilst the bulk energies will be on the order of $T_j$ and $R_j$, the energy distribution of the edge mode is rather different. As mentioned in the main text, the edge mode energy for a particular disorder realization can be estimated as $\log E_\text{maj} \sim \pm \sum_j  \log R_j - \log T_j$ (the sign depends on which phase the physical Hamiltonian is in). Therefore, the distribution of $\log E_\text{maj}$ is approximately Gaussian distributed. In terms of $E_\text{maj}$ itself, we have $P_\text{maj}(E_\text{maj}) \sim 1/E_\text{maj}$ up to logarithmic corrections in $E_\text{maj}$ \cite{Ariel2011}, over an exponentially wide distribution, depending on the distribution of $R_j$ and $T_j$ and the system size.

We are interested in the effect of the edge mode, so to simplify matters we can `freeze out' the bulk mode which yields fast oscillations on a timescale $t \sim E_b^{-1}$. Now we calculate the disorder-average of $S^{(2)}(t)$ as

\begin{myequation}
\bar{S}^{(2)}(t) = \frac{1}{\log(E_+/E_-)}\int_{E_-}^{E_+} \frac{dE}{E}\left(\log 2 - \log\left[1 + \cos^2 E t\right]\right)
\end{myequation}

\noindent where $E_{\pm}$ define the limits of the distribution $P_\text{maj}(E_\text{maj})$. Depending on the order of magnitude of $t$, there are three regimes for this quantity. For $t \ll E_+^{-1}$, $\cos Et \approx 1$ throughout the integral and we get zero. For $t \gg E_-^{-1}$, the term in brackets oscillates rapidly and should be replaced by its average value over a cycle of width $2\pi/t$; this value is $\zeta \coloneqq 2\log(2(2 - \sqrt{2})) = 0.3167\ldots$. In the intermediate regime, the integral should be divided into regions where $E$ is less than or greater than $t^{-1}$. Similar arguments to above then tell us that the former region evaluates to zero, whilst the latter region yields

\begin{myequation}
\bar{S}^{(2)}(t) = \frac{1}{\log(E_+/E_-)}\int_{t^{-1}}^{E_+} \frac{dE}{E} \zeta \propto \log t + \text{const.}
\end{myequation}

From our toy model, we see that disorder averaging over the distribution of edge mode energies yields a logarithmic growth in the entanglement entropy. The limits of the distribution $E_+$ and $E_-$, which depend on the Hamiltonian parameters and the system size, determine the timescales at which the logarithmic growth begins and ends. Additionally, since this phenomenon is due to a single mode, the growth is not unbounded and $\bar{S}^{(2)}(t)$ should increase by an $\mathcal{O}(1)$ amount between times $E_+^{-1}$ and $E_-^{-1}$.

As stated in the main text, to see this logarithmic growth it is crucial that the edge modes in questions are \textit{strong zero modes}, i.e.~operators localized at either edge which commute with the Hamiltonian up to corrections that decay exponentially with system size, but are guaranteed to produce orthogonal states when acting on eigenstates \cite{Fendley2012}. These strong zero modes guarantee that the entire spectrum is nearly degenerate, as opposed to a degeneracy of the ground state only. This is important because the quenches involved generally result in energy densities which are extensive in the system size, and so the initial state has overlap with all the eigenstates of the final Hamiltonian, not just the low-energy eigenstates. Because the slow dynamics will only appear in eigenstates which have near-degeneracies, we conclude that the entire spectrum must be nearly degenerate in order to see the $\log t$ growth. If only the ground state were degenerate, then the magnitude of this growth would be proportional to the overlap of the initial state with the ground state, which is generically exponentially small in the system size.

We also note that the entanglement entropy will only reflect this slow dynamics if the slow mode has non-zero weight on either side of the entanglement cut. For the Majorana mode in question, which is made up of operators on the left and right edges, this criterion is satisfied. However, the log growth would not be seen if, for instance, the slow mode was localized in a generic place in the bulk of the system.

\section{Locality in the physical basis}

Here we prove that the single-particle occupation numbers $\hat{a}_n^\dagger\hat{a}_n$ for systems \eqref{eqHamiltXY} and \eqref{eqHamiltSym} are local when expressed in the physical basis, but that the excitation operators $\hat{a}_n, \hat{a}_n^\dagger$ are non-local. We will consider system \eqref{eqHamiltSym} only, but similar arguments apply for the $XY$ model \eqref{eqHamiltXY}.

We first note that the Hamiltonian \eqref{eqTransTF} contains only terms with one $\hat{\phi}^A$ and one $\hat{\phi}^B$, and also only couples operators with the same Roman numeral $\alpha = \text{I}, \text{II}$. We therefore write

\begin{myequation}
\hat{H}_\text{free} = \sum_{j,k = 1}^N i\hat{\phi}_{j,\text{I}}^A H^{(\text{I})}_{jk} \hat{\phi}_{k,\text{I}}^B + i\hat{\phi}_{j,\text{II}}^A H^{(\text{II})}_{jk} \hat{\phi}_{k,\text{II}}^B
\end{myequation}

\noindent where the matrices $H^{(\text{I})}_{jk}$, $H^{(\text{II})}_{jk}$ are antisymmetric and real. To construct eigenstates (and hence the occupation operators), we use singular value decomposition \cite{Miao2017} to write $H^{(\text{I})} = (U^{(\alpha)})^T \Lambda^{(\alpha)} V^{(\alpha)}$ fo $\alpha = \text{I}, \text{II}$, where $U^{(\alpha)}$ and $V^{(\alpha)}$ are real orthogonal matrices, and $\Lambda^{(\alpha)}$ are diagonal. If we define diagonalized Majorana operators $\hat{\chi}_{n,\alpha}^A = U_{n,j}^{(\alpha)} \hat{\phi}_{j,\alpha}^A$ and $\hat{\chi}_{n,\alpha}^B = V_{n,j}^{(\alpha)} \hat{\phi}_{j,\alpha}^B$ (which satisfy the Majorana commutation relations) then we can write
\begin{myequation}
\hat{H}_\text{free} = \sum_{\alpha = \text{I}, \text{II}} \sum_n i\Lambda^{(\alpha)}_n \hat{\chi}_{n,\alpha}^A \hat{\chi}_{n,\alpha}^B.
\label{eqDiagonalMajorana}
\end{myequation}
We finally write the creation and annihilation operators as $\hat{a}_{n,\alpha} = (\hat{\chi}_{n,\alpha}^A + i\hat{\chi}_{n,\alpha}^B)/2$ and $\hat{a}_{n,\alpha}^\dagger = (\hat{\chi}_{n,\alpha}^A - i\hat{\chi}_{n,\alpha}^B)/2$, from which the single-particle orbitals can be constructed, and the Hamiltonian takes the desired form $\hat{H}_\text{free} = \sum_n \epsilon_n \hat{a}_n^\dagger \hat{a}_n$ (where the $\alpha$ index is suppressed).

When $T_j$ and/or $R_j$ are disordered the free-fermion system \eqref{eqHamiltFree} becomes Anderson localized \cite{Yu2013}. This means that the single-particle eigenstates of the matrices $H^{(\alpha)}$ (i.e.~the rows of $U^{(\alpha)}$ and $V^{(\alpha)}$) decay exponentially away from some site. Thus the occupation numbers $\hat{a}_{n,\alpha}^\dagger \hat{a}_{n,\alpha}$, when expressed in the $\hat{\lambda}$ (equivalently, $\hat{\phi}$) degrees of freedom, are also local. To be precise in what we mean by `local', we can use the definition given in Ref.~\cite{Huse2014}: Let us expand $\hat{a}_{n,\alpha}^\dagger \hat{a}_{n,\alpha}$ in terms of products of on-site Majorana operators. We define the range of each term of the sum as the maximum distance between two sites which are acted on non-trivially by the Majorana product operator. For example, $\hat{\lambda}_{j}^A\hat{\lambda}_{j+r}^B$ has range $|r|$. The range of the operator is the average (weighted by the modulus of the coefficients) of the range of each term, which is finite in a localized system. This range is indeed finite for the free-fermion system, and we must show that the same is true for system \eqref{eqHamiltSym}.

The operators $\hat{a}_{n,\alpha}^\dagger \hat{a}_{n,\alpha}$ are bilinears of $\hat{\phi}$ fermions with the same $\alpha$ index (equal to I or II) and the \textit{opposite} Majorana flavour index $\mu = A$ or $B$. On examining the form of the transformation \eqref{eqFullTrans}, we see that $\hat{\phi}^A_\text{I}$ and $\hat{\phi}^B_\text{I}$ have the same `type' of JW strings (i.e.~identical products of operators from site 1), and the same is true for $\alpha = \text{II}$. This means that when an individual bilinear term $b_{j,\alpha}^n \hat{\phi}_{j,\alpha}^A \hat{\phi}_{k,\alpha}^B$ is transformed into the physical basis, (for $j < k$) a JW string between sites $j$ and $k$ will appear, but the JW strings acting between sites 1 and $j$ will cancel. This cancellation of the JW string means that under the transformation, \textit{the range of any given term remains the same}. Therefore if $\hat{a}_{n,\alpha}^\dagger \hat{a}_{n,\alpha}$ has a finite range in the JW basis, it must also have a finite range in the physical basis. One can also make the same arguments for the $XY$ model \eqref{eqHamiltXY}.

However, the same is not true for the individual excitation operators $\hat{a}_j^\dagger$. Since these are linear in the $\hat{\phi}$ operators, the transformation into the physical basis will yield a non-cancelling JW string for every term. Indeed when $R_j = 0$, an excitation of the $XY$ model is a domain wall at site $j$ separating regions of $\langle \sigma_{k \leq j}^x \rangle = -1$ and $\langle \sigma_{k>j}^x \rangle =+1$. To excite this domain wall from the ferromagnetic ground state requires non-trivial action on all sites from 1 to $j$, which explains why the excitation operators must be non-local.

\end{document}